\documentclass[ conference,twocolumn,twoside]{IEEEtran} 

\usepackage{graphicx}
\usepackage{float}
\usepackage{algorithmic}
\usepackage{array}
\usepackage{amsmath}
\usepackage{amssymb}
\usepackage{mdwmath}
\usepackage{mdwtab}
\usepackage{eqparbox}
\usepackage{stfloats}
\usepackage{fixltx2e}
\usepackage{cases}

\usepackage{xcolor}
\usepackage{makecell}

\usepackage[boxed,ruled,commentsnumbered]{algorithm2e}
\usepackage{url}
\usepackage{cite}
\ifCLASSOPTIONcompsoc
\usepackage[caption=false,font=normalsize,labelfont=sf,textfont=sf]{subfig}
\else
\usepackage[caption=false,font=footnotesize]{subfig}
\fi
\allowdisplaybreaks[4]

\newtheorem{proposition}{Proposition}

\newcommand{\diag}{\mathop{\mathrm{diag}}}
\newcommand{\tr}{\mathop{\mathrm{trace}}}

\makeatletter
\renewcommand*{\@opargbegintheorem}[3]{\trivlist
      \item[\hskip \labelsep{\bfseries #1\ #2}] \mathbf{(#3):}\ }
\makeatother

\begin{document}

\title
{Signal Shaping for Semantic Communication Systems with A Few Message Candidates}
%
\author{\IEEEauthorblockN{  Shuaishuai~Guo$^{*,\dag}$,~
                           Yanghu~Wang$^{*,\dag}$, and
                            Peng~Zhang$^\ddagger$
                            }
    \IEEEauthorblockA{\\$^*$School of Control Science and Engineering,
    Shandong University,
    Jinan, China, 250100\\
$^\dag$Shandong Provincial Key Laboratory of Wireless Communication Technologies, Shandong University,
    Jinan, China, 250100\\
$^\ddagger$School of Computer Engineering, Weifang University, China, 261061\\
    Email: shuaishuai\_guo@sdu.edu.cn, yh-wang@mail.sdu.edu.cn, sduzhangp@163.com
}
}
\maketitle

\pagestyle{empty}
\thispagestyle{empty}

\begin{abstract} Semantic communications target to reliably convey the semantic meaning of messages. It is different from existing communication systems focusing on reliable bit transmission. To achieve the goal of semantic communications, we propose a signal shaping method by minimizing the semantic loss, which is measured by the pretrained bidirectional encoder representation from transformers (BERT) model. The signal set optimization problem for semantic communication systems with a few message candidates is investigated. We propose an efficient projected gradient descent method to solve the problem and prove its convergence.  Simulation results show that the proposed method outperforms existing signal shaping methods in minimizing the semantic loss.

\end{abstract}


\section{Introduction} 
\IEEEPARstart{S}{emantic} communications have been recognized as a promising technology in the sixth-generation (6G) wireless networks since only transmitting the meaning or content of information\cite{Shi2018, Tong2021, Hoydis2021}. However, the difficulty of representing semantic information with precise mathematical models severely limits the development of semantic communications\cite{Qin2022}. Thanks to the recent advances of deep learning (DL), the research on DL based semantic communications has attracted a lot of interest. Most of them realized semantic information transmission by designing the end-to-end communication systems, e.g.,\cite{Farsad2018Deep, Xie2020, Xie2021Lite, Kurka2020Deep, Weng2021Semantic}. To be more specific, Farsad et al. in \cite{Farsad2018Deep} proposed a semantic codec scheme based on bidirectional long short term memory (BLSTM), which achieved lower the word error rate. Xie et al. in\cite{Xie2020} designed a semantic communication system based on Transformer, which not only uses transfer learning to ensure that the system is suitable for different channel environments, but also uses semantic similarity to verify the effectiveness of the system. Further, Xie et al.\cite{Xie2021Lite} proposed a lite distributed semantic communication system based on previous work\cite{Xie2020}, for Internet-of-Things devices with limited computing capability. Besides, Kurka et al. in\cite{Kurka2020Deep} designed a neural network architecture with output feedback based on autoencoder for image source. Weng et al. in\cite{Weng2021Semantic} proposed a  attention mechanism based semantic communication system for speech signals.

So far, DL-enabled semantic communications have shown impressive capabilities, especially at low signal-to-noise (SNR) regime. In fact, however, most of them still aim to secure the bit-level/words-level/message-level precision, e.g.,\cite{Xie2020, Xie2021Lite, Jiang2021}. This is because they use traditional bit-level/words-level/message-level  supervisions, such as cross entropy (CE) or bit/word/message error rate, which are difficult to catch the semantics\cite{Lu2021}. In this letter, we propose a signal shaping method for semantic communication systems with a few message candidates.  We formulate a signal set optimization problem to minimize the semantic loss measured by the pretrained bidirectional encoder representation from transformers (BERT) model\cite{Peters2018}. The formulated problem is transformed to a vector optimization subject to a power constraint and solved by an efficient projected gradient descent method. Simulation results show that the proposed method considerably outperforms existing signal shaping methods, which aim to minimize the error rate, in reducing the semantic loss.

\section{System Model} \label{II}
\begin{figure}
        \centering
        \includegraphics[height=1.5cm,width=1\linewidth]{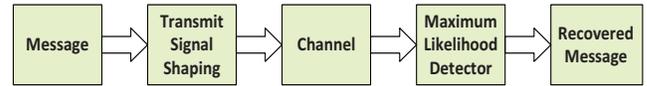}
        \centering\caption{A message semantic communication system}
        \label{System_model}
\end{figure}

\begin{figure}
        \centering
        \includegraphics[height=6.3cm,width=0.9\linewidth]{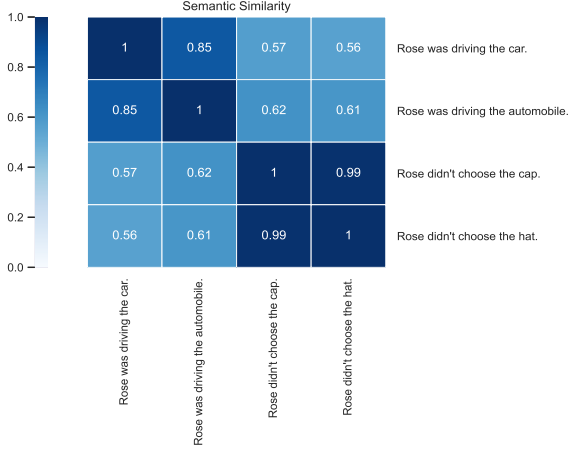}
        \centering\caption{An example to show the semantic similarity computed by the pretrained BERT model.}
        \label{Example}
\end{figure}

This letter considers a message semantic communication system as illustrated in Fig.~\ref{System_model}. In Fig.~\ref{System_model}, a message $\mathbf{m}_i$ chosen from a small message set $\mathcal{M}$ of size $M$. It is then mapped to the $i$th signal vector $\mathbf{x}_i\in\mathbb{C}^{N}$, where $N$ denotes the number of channel uses. Signals carrying messages may be wrongly detected when passing through the noisy channel. Existing communications systems use bit/symbol/message error rate as the performance metrics for the system design, which overlooks the semantic loss of different message error detection. Previously, there has not any efficient way to measure the semantic loss.  Thanks to the advance of deep learning and its applications in nature language processing, some trained models  provide an efficient way to quantify the semantic similarity between two different messages. Based on the pretrained BERT model in \cite{Peters2018}, we define the semantic loss $A(i,j)$ between message $\mathbf{m}_i$  and $\mathbf{m}_j$ as
\begin{equation}
A(i,j)=1-\phi(\mathbf{m}_i,\mathbf{m}_j),
\end{equation}
where $\phi(\mathbf{m}_i,\mathbf{m}_j)=\frac{\mathbf{B_\Phi}(\mathbf{m}_i) \cdot \mathbf{B_\Phi}(\mathbf{m}_j)^T}{\parallel \mathbf{B_\Phi}(\mathbf{m}_i)\parallel \parallel \mathbf{B_\Phi}(\mathbf{m}_j)\parallel }$ denotes the semantic similarity ranging from 0 to 1, and $\mathbf{B_\Phi}(\cdot)$ represents the pretrained BERT model,  which includes billions of parameters and is used for extracting the semantic information. In Fig. \ref{Example}, we show a table listing the semantic similarity of $4$ messages. It can be seen that the value of $\phi(\mathbf{m}_i,\mathbf{m}_j)$ can well reflect the semantic similarity between $\mathbf{m}_i$ and $\mathbf{m}_j$. 

Without loss of generality, the signal vectors are assumed to meet an average normalized power constraint, i.e., $\mathbf{E}(||\mathbf{x}_i||_2^2)\leq 1$. 
Given all messages being transmitted with an equal probability $\frac{1}{M}$, the power constraint can be expressed as $\frac{1}{M}\sum_{i=1}^{M}\mathbf{x}_i^H \mathbf{x}_i\leq 1$. The channel considered in this letter is the additive white Gaussian noise (AWGN) channel that adds noise to achieve a given signal-to-noise ratio (SNR) $\gamma$. Let $\mathcal{X}$ represent the set of all legitimate transmitted signal vectors corresponding to $M$ messages, i.e., $\mathcal{X}=\{\mathbf{x}_1,\mathbf{x}_2,\mathbf{x}_3,\cdots,\mathbf{x}_{M}\}$. At the receiver, the maximum likelihood (ML) detection can be performed by
\begin{equation}
\hat{\mathbf{x}}_i=\arg\max_{\mathbf{x}_i\in\mathcal{X}}\mathbf{p}(\mathbf{y}|\mathbf{x}_i).
\end{equation}
As $\mathbf{p}(\mathbf{y}|\mathbf{x}_i)\propto \exp(-||\mathbf{y}-\mathbf{x}_i||_2^2)$,  the ML detector can be further expressed as
\begin{equation}
\hat{\mathbf{x}}_i=\arg\max_{\mathbf{x}_i\in\mathcal{X}}||\mathbf{y}-\mathbf{x}_i||_2^2.
\end{equation}

\section{Problem Formalization} \label{III}
According to \cite{Tse2005}, the pairwise error detection probability for any $\mathbf{x}_i\neq \mathbf{x}_j$ using the ML detector can be written as
\begin{equation}
P(\mathbf{x}_i,\mathbf{x}_j)=Q\left(\sqrt{\frac{\gamma||\mathbf{x}_i-\mathbf{x}_j||_2^2}{2}}\right)
\end{equation}
where  $Q(x)=\frac{1}{\sqrt{2\pi}}\int_{x}^{\infty}\exp\left(-\frac{u^2}{2}\right)du$.
The corresponding semantic loss caused by the wrong detection between $\mathbf{x}_i$ and $\mathbf{x}_j$ can be expressed by
\begin{equation}\label{eq6}
SL(\mathbf{m}_i,\mathbf{m}_j)=A(i,j)Q\left(\sqrt{\frac{\gamma||\mathbf{x}_i-\mathbf{x}_j||_2^2}{2}}\right).
\end{equation}
For the transmission of all messages, a union upper bound on the average semantic loss can be expressed as
\begin{equation}\label{eq7}
\overline{SL}(\mathcal{X})= \frac{1}{M}\sum_{i=1}^{M}\sum_{j=1,j\neq i}^{M}SL(\mathbf{m}_i,\mathbf{m}_j).
\end{equation}
Substituting (\ref{eq6}) into (\ref{eq7}), we can compute the upper bound for the semantic loss as
\begin{equation}
\overline{SL}(\mathcal{X})= \frac{1}{M}\sum_{i=1}^{M}\sum_{j=1,j\neq i}^{M}A(i,j)Q\left(\sqrt{\frac{||\gamma\mathbf{x}_i-\mathbf{x}_j||_2^2}{2}}\right).
\end{equation}
Therefore, the optimization problem of signal shaping to reduce the semantic loss can be formulated as
\begin{subequations}
\begin{align}
(\textbf{P1}):~\mathrm{Given}: &~\gamma, A(i,j),\forall i,j\notag\\
\mathrm{Find}:&~\mathcal{X}=\{\mathbf{x}_1,\mathbf{x}_2,\cdots,\mathbf{x}_{M}\}\notag\\
\mathrm{Minimize}:&~\overline{SL}(\mathcal{X})\notag\\
\mathrm{Subject~to}:&~\frac{1}{M}\sum_{i=1}^{M}\mathbf{x}_i^H \mathbf{x}_i\leq 1.\notag
\end{align}
\end{subequations}

\section{The Proposed Signal Shaping Method}\label{IV}
Problem (\textbf{P1}) is a set optimization problem, which is difficult to solve. To make it tractable, we first transform it as a vector optimization problem and then propose an efficient projected gradient descent algorithm to deal with it.
\subsection{Problem Transformation}
First, we rewrite the power of the Euclidean distance between signal vectors as
\begin{equation}
||\mathbf{x}_i-\mathbf{x}_j||_2^2=||\mathbf{G}\mathbf{D}_{\mathbf{z}}(\mathbf{e}_i-\mathbf{e}_j)||_2^2,
\end{equation} 
where 
\begin{equation*}
\mathbf{G}=\overbrace{\left[\mathbf{I}_{N},\mathbf{I}_{N},\cdots,\mathbf{I}_N\right]}^{M}\in \mathbb{C}^{N\times MN},
\end{equation*}
\begin{equation*}
\mathbf{D}_{\mathbf{z}}=\diag (\mathbf{z})\in \mathbb{C}^{MN\times MN},
\end{equation*}
\begin{equation*}
\mathbf{z}=[\mathbf{x}_1^T,\mathbf{x}_2^T,\cdots,\mathbf{x}_{M}^T]^T\in\mathbb{C}^{MN\times 1},
\end{equation*}
\begin{equation*}
\mathbf{e}_i=\mathbf{g}_i\otimes\mathbf{x}_i\in \mathbb{C}^{MN\times1},~\mathbf{e}_j=\mathbf{g}_j\otimes\mathbf{x}_j\in \mathbb{C}^{MN\times1},
\end{equation*}
and $\mathbf{g}_i\in \mathbb{C}^{M\times1}$, $\mathbf{g}_i\in \mathbb{C}^{M\times1}$ represent the $i$th and $j$th one-hot vectors with all zeros except a one at the $i$th position and $j$th position, respectively.
Expanding the expression, we have
\begin{equation}
\begin{split}
||\mathbf{G}\mathbf{D}_{\mathbf{z}}(\mathbf{e}_i-\mathbf{e}_j)||_2^2&=(\mathbf{e}_i-\mathbf{e}_j)^H\mathbf{D}_{\mathbf{z}}^H\mathbf{G}^H\mathbf{G}\mathbf{D}_{\mathbf{z}}(\mathbf{e}_i-\mathbf{e}_j)\\
&=\tr\left(\mathbf{D}_{\mathbf{z}}^H\mathbf{R}_{\mathbf{G}}\mathbf{D}_{\mathbf{z}}\Delta\mathbf{E}_{i,j}\right),
\end{split}
\end{equation}
where $\mathbf{R}_{\mathbf{G}}=\mathbf{G}^H\mathbf{G}$ and $\mathbf{E}_{ij}=(\mathbf{e}_i-\mathbf{e}_j)(\mathbf{e}_i-\mathbf{e}_j)^H$.

Summarizing above and based on the rule $\tr(\mathbf{D}_{\mathbf{u}}\mathbf{A}\mathbf{D}_{\mathbf{v}}\mathbf{B})=\mathbf{u}^H(\mathbf{A}\odot \mathbf{B})\mathbf{v}$ \cite{Zhang2017},
we have
\begin{equation}
||\mathbf{x}_i-\mathbf{x}_j||_2^2=\mathbf{z}^H\mathbf{W}_{ij}\mathbf{z},
\end{equation}
where $\mathbf{W}_{ij}=\mathbf{R}_{\mathbf{G}}\odot\Delta\mathbf{E}_{ij}$.

With such transformation, the upper bound for the semantic loss can be rewritten as
\begin{equation}
\overline{SL}(\mathbf{z})=\frac{1}{M}\sum_{i=1}^{M}\sum_{j=1,j\neq i}^{M}A(i,j)Q\left(\sqrt{\frac{\gamma\mathbf{z}^H\mathbf{W}_{ij}\mathbf{z}}{2}}\right),
\end{equation}
and the power constraint can be rewritten as
\begin{equation}
\mathbf{z}^H\mathbf{z}\leq M.
\end{equation}

Therefore, the set optimization problem can be reformulated as a vector optimization problem:
\begin{subequations}
\begin{align}
(\textbf{P2}):~\mathrm{Given}: &~\gamma,\mathbf{W}_{ij}, A(i,j),\forall i,j\notag\\
\mathrm{Find}:&~\mathbf{z}\notag\\
\mathrm{Minimize}:&~\overline{SL}(\mathbf{z})\notag\\
\mathrm{Subject~to}:&~\mathbf{z}^H\mathbf{z}\leq M. \notag
\end{align}
\end{subequations}
\subsection{Projected Gradient Descent Optimization Method}
Problem (\textbf{P2}) is a non-convex problem and the optimal solution to the problem (\textbf{P2}) is not unique\footnote{It is obvious that a same phase rotation on all signal vectors will not change their mutual Euclidean distances. Thus, a phase rotation on the optimal solution is still an optimal solution to achieve the minimum semantic loss.}. To solve the problem (\textbf{P2}), we formulate the Lagrangian function as
\begin{equation}
L(\mathbf{z},\lambda)=\overline{SL}(\mathbf{z})+\lambda(\mathbf{z}^H\mathbf{z}-M).
\end{equation}
According to the Karush-Kuhn-Tucker (KKT) conditions, the optimal solutions to the problem (\textbf{P2}) should satisfy
\begin{equation}\label{KKT}
\left\{
\begin{aligned}
&\nabla_{\mathbf{z}}L(\mathbf{z},\lambda)=0,\\
&\lambda(\mathbf{z}^H\mathbf{z}-M)=0,\\
&\lambda\geq 0.
\end{aligned}
\right.
\end{equation}
Because $L(\mathbf{z},\lambda)$ monotonically decrease with the power, thus it is minimized when the power constraint is met with strict equality, i.e.,
$\lambda(\mathbf{z}^H\mathbf{z}-M)=0$. The first condition in (\ref{KKT}) can be expressed as
\begin{equation}\label{KKT1}
\nabla_{\mathbf{z}}L(\mathbf{z},\lambda)=\left[\Omega(\mathbf{z})+2\mu \mathbf{I}_{MN}\right]\mathbf{z}=0,
\end{equation}
where 
\begin{equation}\label{eq18}
\Omega(\mathbf{z})=-\frac{1}{M}\sum_{i=1}^{M}\sum_{j=1,j\neq i}^{M}\sqrt{\frac{\gamma A(i,j)^2}{{4\pi\mathbf{z}^H\mathbf{W}_{ij}}\mathbf{z}}}\cdot e^{-\frac{\gamma\mathbf{z}^H\mathbf{W}_{ij}\mathbf{z}}{4}}\cdot \mathbf{W}_{ij}.
\end{equation}
Clearly, the closed-form solution to (\ref{KKT1}) is difficult to obtain. In this letter, we resort to a projected gradient descent method to find a good solution. In detail, we first compute the gradient descent direction  in the $k$th iteration as
\begin{equation}\label{eq21}
\mathbf{g}_{k}=-\Omega(\mathbf{z}_k)\mathbf{z}_k.
\end{equation}
To ensure the power of the solution unchanged, we perform a projection by
\begin{equation}\label{eq22}
\mathbf{g}_k^{\bot}=\mathbf{g}_k-\frac{\mathbf{z}_k^H\mathbf{g}_k\mathbf{z}_k}{||\mathbf{z}_k||_2^2},
\end{equation}
such that $\mathbf{z}_k^H\mathbf{g}_k^{\bot}=0$.
After projection, we update the solution to be
\begin{equation}\label{eq23}
\mathbf{z}_{k+1}=\cos \theta\cdot \mathbf{z}_k+\sin \theta \cdot\sqrt{M}\frac{\mathbf{g}_k^{\bot}}{\left\|\mathbf{g}_k^{\bot}\right\|_2},
\end{equation}
where $\theta\in[0,\frac{\pi}{2}]$ can be obtained by solving
\begin{equation}\label{eq24}
\theta=\arg\min_{\theta\in[0,\frac{\pi}{2}]} \overline{SL}(\mathbf{z}_{k+1}).
\end{equation}
By updating the solution until the stop criterion $\frac{\|\mathbf{g}_k^{\bot}\|_2}{\|\mathbf{g}_k\|_2}\leq \epsilon$ is met, we can obtain a good solution. 
For clarity, we summarize the projected gradient descent algorithm in Algorithm 1.

\begin{algorithm}[t] 
\caption{Projected gradient descent algorithm for minimizing semantic loss in semantic communications}
\label{alg:B}
\begin{algorithmic}
\STATE Initialize $k=1$, $\epsilon$, and $\mathbf{z}_1$ with $\mathbf{z}_1^H\mathbf{z}_1=M$.
\REPEAT
\STATE Compute the gradient descent direction by (\ref{eq21}).
\STATE Perform projection by (\ref{eq22}).
\STATE Search $\theta$ by solving (\ref{eq24}).
\STATE Update $\mathbf{z}_{k+1}$ by (\ref{eq23}).
\STATE $k\leftarrow k+1$
\UNTIL{$\frac{\|\mathbf{g}_k^{\bot}\|_2}{\|\mathbf{g}_k\|_2}\leq \epsilon$}
\STATE Output $\mathbf{z}=\mathbf{z}_k$.
\end{algorithmic}
\end{algorithm}

\subsection{Convergence and Computational Complexity Analysis}
Since the semantic loss is lower bounded, Algorithm 1 converges because of the following proposition.
\begin{proposition}
 Algorithm 1 always guarantees $\overline{SL}(\mathbf{z}_{k+1})\leq \overline{SL}(\mathbf{z}_{k})$.
\end{proposition}
\begin{IEEEproof}
For a sufficient small $\theta$ with $\theta\rightarrow 0$, we can derive the first-order Taylor expansion of $\overline{SL}(\mathbf{z}_{k+1})$
based on (\ref{eq23}) as  
\begin{equation}
\overline{SL}(\mathbf{z}_{k+1})\approx\overline{SL}(\mathbf{z}_k)-\mathbf{g}_k^H\cdot\sqrt{M}\frac{\mathbf{g}_k^{\bot}}{\left\|\mathbf{g}_k^{\bot}\right\|_2}\cdot\theta.
\end{equation}
Since $\mathbf{g}_k^H\mathbf{g}_k^{\bot}=(1-\cos^2\alpha)||\mathbf{g}_k||_2^2\geq 0$, where $\alpha=\arccos \frac{<\mathbf{g}_k,\mathbf{z}_k>}{||\mathbf{g}_k||_2||\mathbf{z}_k||_2}$, therefore we have
\begin{equation}
\overline{SL}(\mathbf{z}_{k+1})\leq \overline{SL}(\mathbf{z}_{k}).
\end{equation}
\end{IEEEproof}

The computational complexity of Algorithm 1 mainly comes from the computation of the gradient descent direction by (\ref{eq18}) and (\ref{eq21}). It can be analyzed to be $\mathcal{O}(N_{iter}M^4N^2)$, where $N_{iter}$ represents the number of iterations that Algorithm 1 performs. Even though the computational complexity is high for large $M$, it will not be a problem as the signal shaping can be conducted offline.
\setcounter{figure}{3}
 \begin{figure*}
        \centering
        \includegraphics[height=6.5cm,width=1\linewidth]{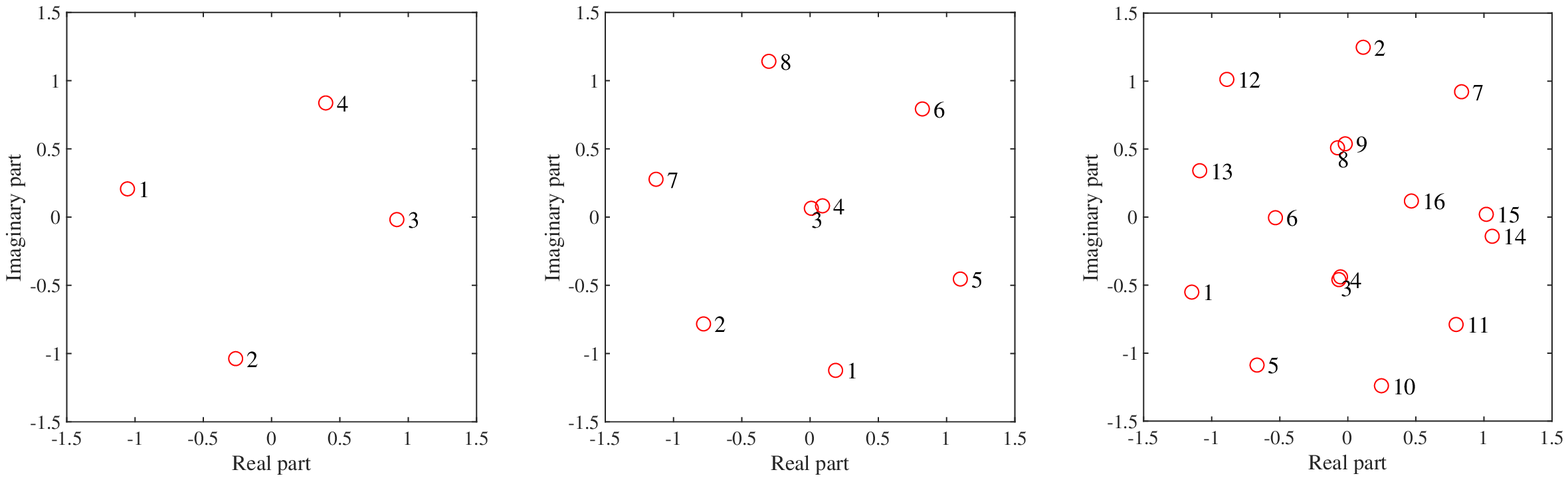}
\centering\caption{Signal constellation designs for semantic communication systems with $M=4$, $8$ and $16$ messages and $N=1$ at $\gamma=10~\mathrm{dB}$.}
        \label{result2}
\end{figure*}
\section{Simulation and Discussion}\label{V}
In this section, we first investigate the convergence of the proposed algorithm for signal shaping and its sensitivity to the randomly generated initial solution $\mathbf{z}_1$ by simulations. In the simulations, we set $M=16$, $N=1$, $\gamma=10~\mathrm{dB}$ and $\epsilon=10^{-2}$. Algorithm 1 is performed $10$ times with $10$ different initial solutions in the simulations. The results are demonstrated in Fig. \ref{result1}. It is shown that
Algorithm 1 converge fast and is sensitive to the initial solution. To combat this sensitivity, we optimize the signal shaping with $50$ randomly generated initial solutions and choose the one with the minimum semantic loss in the following simulations.
\setcounter{figure}{2}
 \begin{figure}
        \centering
        \includegraphics[height=7cm,width=1\linewidth]{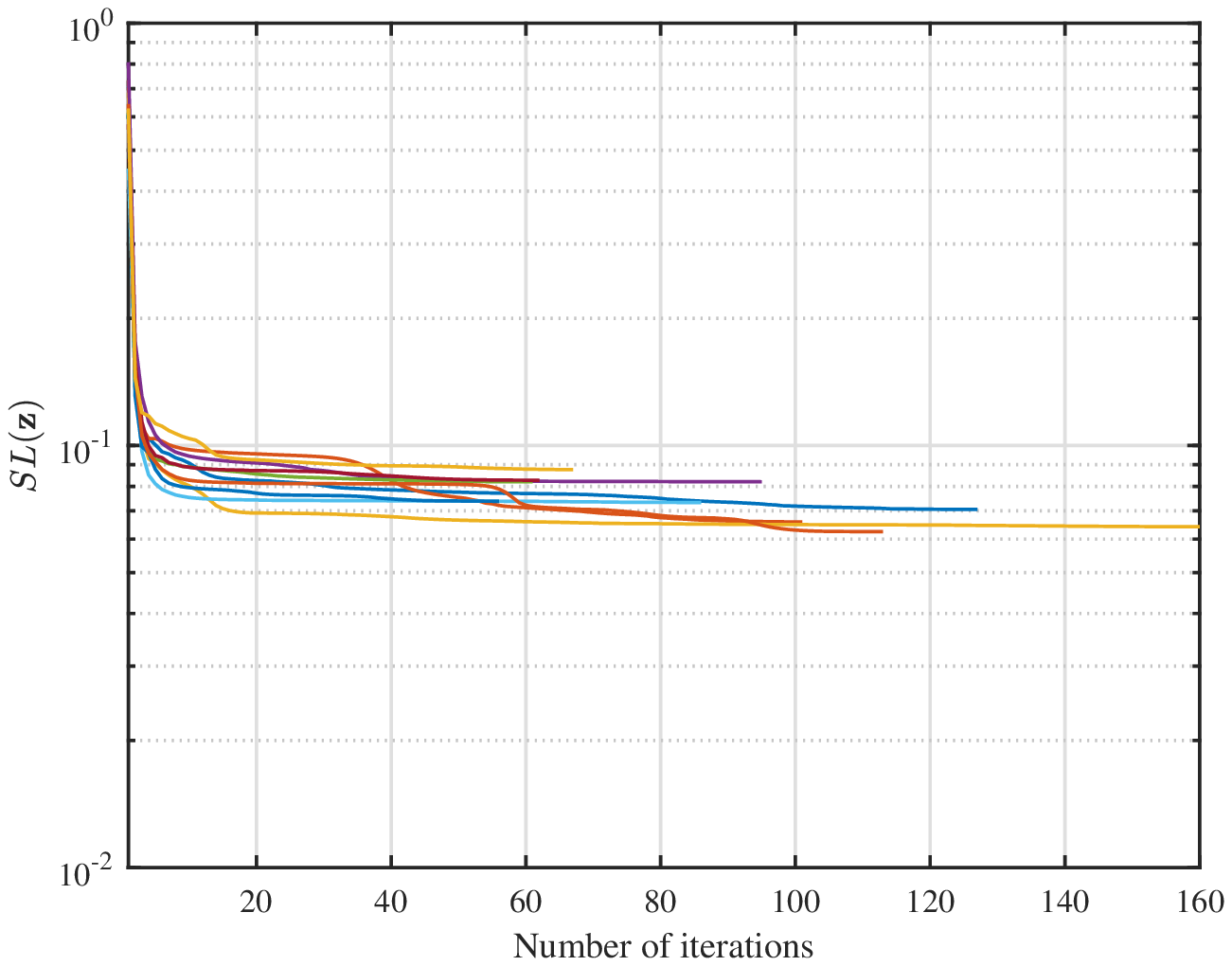}
        \centering\caption{Convergence property of Algorithm 1 with $10$ different randomly generated solutions.}
        \label{result1}
\end{figure}

\setcounter{figure}{4}
 \begin{figure}
        \centering
        \includegraphics[height=7cm,width=1\linewidth]{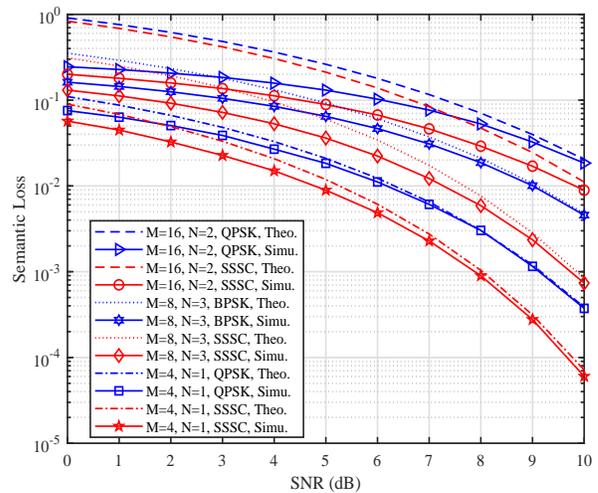}
        \centering\caption{Semantic loss of different signal shaping methods.}
        \label{result3}
\end{figure}

Second, we show the optimized signal designs for message semantic communication systems with $M=4,8,16$ candidate messages\footnote{The messages to be delivered and their semantic similarity matrices are available at \url{https://github.com/SSG-SDU/Semantic-Similarity/tree/master}.}. Other parameters $N$, $\epsilon$, $\gamma$ are set as $1$, $10^{-2}$, and $10 \mathrm{dB}$, respectively.  The signal designs are illustrated in Fig. \ref{result2}. It is shown that the designed signal constellations are irregular, which are different from traditional ones. This is because the semantic meaning of each signal points are taken into considerations. Those signal constellation points have larger semantic similarities are closer to each other, leading to that more space is left to place other signal constellation points.

Third, to show the superiority of the proposed signal shaping for semantic communications (SSSC), we compare it with those signal designs that can achieve the minimum message error rate, e.g., binary phase shift keying (BPSK) and quadrature phase shift keying (QPSK). Simulation results are demonstrated in Fig. \ref{result3}. Simulation results show that taking the semantic meaning of messages into consideration can bring considerable performance gain in reducing the semantic loss in semantic communication systems. Specifically, under the setup with $M=4,~N=1$, SSSC is better than QPSK by $1$ $\mathrm{dB}$ at the  semantic loss of $10^{-3}$. Under the setup with $M=8,~N=3$, SSSC is better than BPSK by around $0.8$ $\mathrm{dB}$ at the semantic loss of $10^{-2}$. Under the setup with $M=16,~N=2$, SSSC is better than QPSK by $1.3$ $\mathrm{dB}$ at the semantic loss of $2\times 10^{-2}$. Besides, we include the numerical results of the theoretical upper bound of semantic loss and it is shown that the theoretical upper bound is tight at high SNR regime.

\section{Conclusion}
In this letter, we proposed a signal shaping method to minimize the semantic loss for semantic communication systems with a few message candidates. The semantic loss was quantified by the pretrained BERT model. We proposed an efficient projected gradient descent method to deal with the problem. We compared the proposed signal design with existing signal designs that achieve the minimum message error rate. Simulation results demonstrated that the proposed signal shaping method can provide considerable gain in reducing the semantic loss. 


\end{document}